\documentclass[a4paper]{article}
\usepackage{RR}
\RRdate{January 2007}

\newtheorem{lemma}{Lemma}
\newtheorem{corollary}{Corollary}
\newtheorem{theorem}{Theorem}

\newenvironment{proof}{\textbf{Proof:}}{$\hfill\Box$}

\usepackage{amssymb,graphicx,amsmath}
\usepackage{algorithm}
\usepackage[noend]{algpseudocode}

\RRtitle{Un Nouvel Algorithme Auto-stabilisant pour le Mariage Maximal}
\RRetitle{A New Self-Stabilizing Maximal Matching Algorithm}

\titlehead{Self-stabilizing Maximal Matching}

\RRauthor{Fredrik Manne\thanks[bergen]{University of Bergen, Norway, {\tt \{fredrikm,mortenm\}@ii.uib.no}} \and Morten Mjelde\thanksref{bergen} \and Laurence Pilard\thanks{University of Iowa, USA, {\tt laurence-pilard@uiowa.edu}} \and S\'{e}bastien Tixeuil\thanks{LRI-CNRS UMR 8623 \& INRIA Grand Large,
Universit\'{e} Paris Sud, France, {\tt tixeuil@lri.fr}}}

\RRresume{Le probl\`{e}me du mariage maximal a re\c cu une attention consid\'{e}rable de la part de la communaut\'{e} de l'auto-stabilisation. Les travaux pr\'{e}c\'{e}dents ont propos\'{e} des algorithmes auto-stabilisants pour r\'{e}soudre le probl\`{e}me \`{a} la fois dans le cas d'un d\'{e}mon non-\'{e}quitable, \'{e}quitable, distribu\'{e}, s\'{e}quentiel, et synchrone. Dans cet article, nous pr\'{e}sentons un algorithme auto-stabilisant pour ce probl\`{e}me qui unifie toutes les approches pr\'{e}c\'{e}dentes au sens o\`{u} sa complexit\'{e} en nombre de pas de calcul est identique \`{a} la meilleure complexit\'{e} connue pour les d\'{e}mons s\'{e}quentiel non-\'{e}quitable, distribu\'{e} \'{e}quitable, et synchrone. En outre, l'algorithm am\'{e}liore la meilleure complexit\'{e} connue pour le d\'{e}mon non-\'{e}quitable distribu\'{e} de $O(n^2)$ et $O(\delta n)$ \`{a} $O(m)$, o\`{u} $n$ est le nombre de processus, $m$ le nombre d'ar\^{e}tes, et $\delta$ le degr\'{e} maximum du graphe.} 

\RRabstract{The maximal matching problem has received considerable attention in the self-stabilizing
community. Previous work has given different self-stabilizing algorithms that solves the problem for
both the adversarial and fair distributed daemon, the sequential adversarial daemon, as well as the
synchronous daemon.
In the following we present a single self-stabilizing algorithm for this problem that unites all of these
algorithms in that it stabilizes in the same number of moves as the previous best algorithms for the
sequential adversarial, the distributed fair, and the synchronous daemon. In addition, the algorithm
improves the previous best moves complexities for the distributed adversarial daemon from $O(n^2)$ and
$O(\delta m)$ to $O(m)$ where $n$ is the number of processes, $m$ is the number of edges, and $\delta$ is the maximum
degree in the graph.
}

\RRmotcle{Syst\`{e}mes distribu\'{e}s, Algorithme distribu\'{e}, Auto-stabilisation, Mariage maximal, Complexit\'{e}}
\RRkeyword{Distributed systems, Distributed algorithm, Self-stabilization, Maximal matching, Complexity}

\RRprojet{Grand large}
\RRtheme{\THNum}
\URFuturs

\begin{document}
\makeRR

\section{Introduction}

A \emph{matching} in an undirected graph is a subset of edges in which no pair of edges is adjacent. A matching $M$ is
\emph{maximal} if no proper superset of $M$ is also a matching. Matchings are typically used in distributed applications when pairs of neighboring nodes have to be set up (\emph{e.g.} between a server and a client). As current distributed applications usually run continuously, it is expected that the system is dynamic (nodes may leave or join the network), so an algorithm for the distributed construction of maximal matching should be able to reconfigure on the fly.
\emph{Self-stabilization}~\cite{Dij74,Dol00} is an elegant approach to forward recovery from transient faults as well as initializing a large-scale system. Informally, a self-stabilizing systems is able to recover from any transient fault in finite time, without restricting the nature or the span of those faults.

The environment of a self-stabilizing algorithm is modeled by the notion of \emph{daemon}. There are two main characteristics
for the daemon: it can be either \emph{sequential} (or central, meaning that exactly one eligible process is scheduled for
execution at a given time) or \emph{distributed} (meaning that any subset of eligible processes can be scheduled for
execution at a given time), and in an orthogonal way, it can be \emph{fair} (meaning that in any execution, every eligible
processor is eventually scheduled for execution) or \emph{adversarial} (meaning that the daemon only guarantees global
progress, \emph{i.e.} some eligible process is eventually scheduled for execution). An extreme case of a fair daemon is the
\emph{synchronous} daemon, where all eligible processes are scheduled for execution at every step. Of course, any algorithm that can
cope with the distributed daemon can cope with the sequential daemon or the synchronous daemon, and any algorithm that can
handle the adversarial daemon can be used with a fair or a synchronous daemon, but the converse is not true in either case.

There exists several self-stabilizing algorithms for computing a maximal matching in an unweighted general graph.
Hsu and Huang \cite{HH92} gave the first such algorithm and proved a bound of $O(n^3)$ on the number of steps assuming an
adversarial daemon. This analysis was later improved to $O(n^2)$ by Tel \cite{Tel94b} and finally to $O(m)$ by Hedetniemi et
al.~\cite{HJS01}. The original algorithm assumes an anonymous networks and operates therefore under the sequential daemon in
order to achieve symmetry breaking.

By using randomization, Gradinariu and Johnen~\cite{GJ01} provide a scheme to give processes a local name that is unique within distance $2$, and use this scheme to run Hsu and Huang's algorithm under an adversarial distributed daemon. However, only a finite stabilization time is proved. Using the same technique of randomized local symmetry breaking, Chattopadhyay et al.~\cite{CHS02} later provide a maximal matching with $O(n)$ round complexity (in their model, this is tantamount to $O(n^2)$ steps), but assuming the weaker fair distributed daemon. 

In \cite{GHJ03} Goddard \emph{et al.}~describe a synchronous version of Hsu and Huang's algorithm and show that it stabilizes
in $O(n)$ rounds. Although not explicitly proved in the paper, it can be shown that their algorithm also copes with the
adversarial distributed daemon using $\theta(n^2)$ steps. Here, symmetry is broken using unique identifiers at every process.
In~\cite{GT06}, Gradinariu and Tixeuil provide a general scheme to transform an algorithm using the sequential adversarial
daemon into an algorithm that copes with the distributed adversarial daemon. Using this scheme with Hsu and Huang's algorithm
yields a step complexity of $O(\delta m)$, where $\delta$ denotes the maximum degree of the network.

Our contribution is a new self-stabilizing algorithm that stabilizes after $O(m)$ steps both under the sequential and under
the distributed adversarial daemon. Under a distributed fair daemon the algorithm stabilizes after $O(n)$ rounds.
Thus this algorithm unifies the moves complexities of the previous best algorithms both for the sequential and for the
distributed fair daemon and also improves the previous best moves complexity for the distributed adversarial daemon. 
As a side effect, we
improve the best known algorithm for the adversarial daemon by lowering the environment requirements 
(distributed \emph{vs.} sequential).
To break symmetry, we assume that node identifiers are unique within distance $2$ (this can be done using the scheme of~\cite{GJ01,CHS02}). The following table compares features of aforementioned algorithms and our (best features for each category is presented in boldface).

\medskip

\begin{center}

{\small
\begin{tabular}{|l|l|l|l|} \hline
\textbf{Reference} & \textbf{Daemon} & \textbf{Step complexity} & \textbf{Round complexity} \\ \hline
\cite{HH92,Tel94b,HJS01} & sequential adversarial & $\mathbf{O(m)}$ & \\ 
\cite{GJ01} & \textbf{distributed adversarial} & finite & \\
\cite{CHS02} & distributed fair & O$(n^2)$ & $\mathbf{O(n)}$ \\
\cite{GHJ03} & synchronous & $O(n^2)$ & $\mathbf{O(n)}$ \\
\cite{GT06} & \textbf{distributed adversarial} & $O(\delta m)$ & \\
\textbf{This paper} & \textbf{distributed adversarial} & $\mathbf{O(m)}$ & $\mathbf{O(n)}$ \\ \hline
\end{tabular}
}

\end{center}

\medskip

The rest of this paper is organized as follows. In Section \ref{sec:self} we give a short introduction to self-stabilizing algorithms and the computational environment we use. In Section \ref{sec:algorithm} we describe our algorithm and prove its correctness and speed of convergence in Section \ref{sec:proof}. Finally, in Section \ref{sec:conclusion} we conclude.

\section{Model}
\label{sec:self}

A system consists of a set of processes where two adjacent processes can communicate with each other. 
The communication relation is typically represented by a graph $G=(V,E)$ where each process is represented by a node in $V$
and two processes $i$ and $j$ are adjacent if and only if  $(i,j) \in E$.
The set of neighbors of a node $i \in V$ is denoted by $N(i)$. The neighbors of a set of processes $A \subseteq V$ is defined
as follows $N(A) = \{ j \in V-A, \exists i \in A$ s.t.~$(i,j) \in E\}$. 
A process maintains a set of variables. Each variable ranges over a fixed domain of values. An action has the form $\langle
name \rangle : \langle guard \rangle \longrightarrow \langle command \rangle$. A \emph{guard} is a boolean predicate over the
variables of both the process and those of its neighbors. A \emph{command} is a sequence of statements assigning new values to the variables of the process. 

 A \emph{configuration} of the system is the assignment of a value to every variable of each process from its corresponding
 domain. Each process contains a set of actions.  An action is \emph{enabled} in some configuration if its guard is
 \textbf{true} at this configuration. A process is \emph{eligible} if it has at least one enabled action. A \emph{computation} is a maximal sequence of configurations such that for each configuration $s_i$, the next configuration $s_{i+1}$ is obtained by executing the command of at least one action that is enabled in $s_i$ (a process that executes such an action makes a \emph{move} or a \emph{step}). Maximality of a computation means that the computation is infinite or it terminates in a configuration where none of the actions are enabled. 

A \emph{daemon} is a predicate on executions. We distinguish several kinds of daemons: the \emph{sequential} daemon make the system move from one configuration to the next by executing exactly one enabled action, the \emph{synchronous} daemon makes the system move from one configuration to the next one by executing all enabled actions, the \emph{distributed} daemon makes the system move from one configuration to the next one by executing any non empty subset of enabled actions. Note that the sequential and synchronous daemons are instances of the more general (\emph{i.e.} less constrained) distributed daemon. Also, a daemon is \emph{fair} if any action that is continuously enabled is eventually executed, and \emph{adversarial} if it may execute \emph{any} enabled action at every step. Again, the adversarial daemon is more general than the fair daemon.

A system is self-stabilizing for a given specification, if it automatically converges to a configuration that conforms to this specification, independently of its initial configuration and without external intervention.

We consider two measures for evaluating complexity of self-stabilizing programs. The \emph{step} complexity investigates the maximum number of process moves that are needed to reach a configuration that conforms to the specification (\emph{i.e.} a \emph{legitimate} configuration), for all possible starting configurations. The \emph{round} complexity considers that executions are observed in rounds: a round is the smallest sequence of an execution in which every process that was eligible at the beginning of the round either makes a move or has its guard(s) disabled since the beginning of the round.

\section{The Algorithm}
\label{sec:algorithm}
In the following we present and motivate our algorithm for computing a
maximal matching.  The algorithm is self-stabilizing and does not make
any assumptions on the network topology. A set of edges $M \subseteq
E$ is a {\em matching} if and only if $x,y \in M$ implies that $x$ and
$y$ does not share a common end point.  A matching $M$ is {\em
  maximal} if no proper superset of $M$ is also a matching.

Each process $i$ has a variable $p_i$ pointing to one of its neighbors or to {\em null}.
We say that processes $i$ and $j$ {\em are married} to each other if and only if 
$i$ and $j$ are neighbors and their $p$-values point to each other. In this case we will 
also refer to $i$ as being married without specifying $j$. However, we note that
in this case $j$ is unique. A process which is not married is {\em unmarried}.

We also use a variable $m_i$ to let neighboring processes of $i$ know if process
$i$ is married or not. To determine the value of $m_i$ we use a predicate
{\em PRmarried(i)} which evaluates to true if and only if $i$ is married.
Thus predicate {\em PRmarried(i)} allows process $i$ to know if it is currently
married and the variable $m_i$ allows neighbors of $i$ to know if $i$ is married.
Note that the value of $m_i$ is not necessarily equal to {\em PRmarried(i)}.

Our self-stabilizing scheme is given in Algorithm \ref{alg:matching}.
It is composed of four mutual exclusive guarded rules as described below.
\begin{algorithm}[t]
\caption{A self-stabilizing maximal matching algorithm}
\label{alg:matching}
\vspace{2mm}
\begin{tabbing}
xxx \=xxx \=xxx \=xxx \=xxx \=xxx \=xxx \=xxx \=xxx \=xxx \=xxx \kill
\>{\bf Variables of process $i$:} \\
\> \>$m_i \in$ \{true, false\} \\
\> \>$p_i \in \{null\} \cup N(i)$ \\ \\

\>{\bf Predicate:} \\
\> \>$PRmarried(i) \equiv \exists j \in N(i) : (p_i=j$ {\bf and } $p_j=i)$ \\ \\

\>{\bf Rules:} \\ 
\> \>{\em Update:} \\
\> \> \>{\bf if} $m_i \not = PRmarried(i)$ \\
\> \> \>{\bf then} $m_i := PRmarried(i)$ \\ \\

\> \>{\em Marriage:} \\
\> \> \>{\bf if} $m_i = PRmarried(i)$ {\bf and} $p_i = null$ {\bf and} $\exists
j \in N(i) : p_j=i$\\
\> \> \>{\bf then} $p_i := j$ \\ \\

\> \>{\em Seduction:} \\
\> \> \>{\bf if} $m_i = PRmarried(i)$ {\bf and} $p_i = null$ {\bf and} $\forall
k \in N(i) : p_k \not = i$ \\
\> \> \> \>{\bf and} $\exists j \in N(i) : (p_j = null$ {\bf and}
$j>i$ {\bf and} $\neg m_j)$\\
\> \> \>{\bf then} $p_i := Max \{j \in N(i) : (p_j = null$ {\bf and} $j>i$ {\bf and}
                           $\neg m_j)\}$ \\ \\

\> \>{\em Abandonment:} \\
\> \> \>{\bf if} $m_i = PRmarried(i)$ {\bf and} $p_i = j$ {\bf and} $p_j \not = i$
{\bf and} $(m_j$ {\bf or} $j \leq i)$ \\
\> \> \>{\bf then} $p_i := null$

\end{tabbing}
\end{algorithm}



The {\em Update}
rule updates the value of $m_i$ if it is necessary, while the three other rules can
only be executed if the value of $m_i$ is correct.
In the {\em Marriage} rule, an unmarried process that is currently being pointed
to by a neighbor $j$ tries to marry $j$ by setting $p_i = j$.
In the {\em Seduction} rule, an unmarried process that is not being pointed
to by any neighbor, point to an unmarried neighbor with the objective of marriage. 
Note that the identifier
of the chosen neighbor has to be larger than that of the current process.
This is enforced to avoid the creation of cycles of pointer values.
In the {\em Abandonment} rule, a process $i$ resets its $p_i$ value to {\em null}.
This is done if the process $j$ which it is pointing to does not point back at $i$ and
if either {\em (i)} $j$ is married, or {\em (ii)} $j$ has a lower identifier than
$i$.
Condition {\em (i)} allows a process to stop waiting for an already married process
while the purpose of Condition {\em (ii)} is to break a possible initial cycle of 
$p$-values.

We note that if {\em PRmarried(i)} holds at some point of time then from then
on it will remain true throughout the execution of the algorithm.
Moreover, the algorithm will never actively create a cycle of pointing values 
since the {\em Seduction} rule enforces that $j>i$ before process $i$ will point
to process $j$. Also, all initial cycles are eventually broken since the guard
of the {\em Abandonment} rule requires that $j \leq i$.



\begin{figure}
\includegraphics[width=1.0\textwidth]{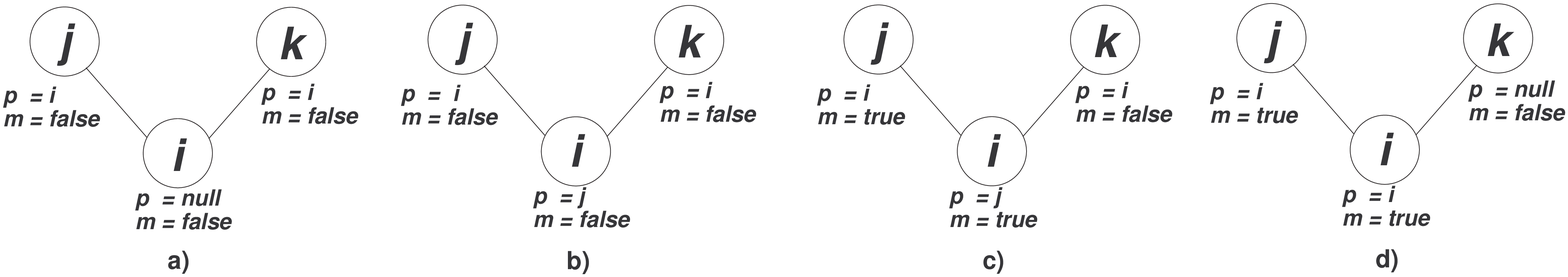}
\caption{\label{fig:fig1} Example}
\end{figure}

Figure \ref{fig:fig1} gives a short example of the execution of the algorithm.
The initial configuration is as shown in Figure \ref{fig:fig1}a, where $id_i > 
id_j > id_k$. Here both processes $j$ and $k$ attempt to 
become married to $i$. 
In Figure \ref{fig:fig1}b process $i$ has executed a {\em Marriage} 
move, and $i$ and $j$ are now married. In Figure \ref{fig:fig1}c both 
$i$ and $j$ execute an {\em Update} move, setting their $m$-values to 
$true$. And finally, in Figure \ref{fig:fig1}d process $k$
executes an {\em Abandonment} move.

\section{Proof of Correctness}
\label{sec:proof}
In the following we will first show that when Algorithm \ref{alg:matching} has reached
a stable configuration it also defines a maximal matching. We will
then bound the number of steps the algorithm needs to stabilize both for the adversarial and fair distributed
daemon. 
Note that the sequential daemon is a 
subset of the distributed one, thus any result for the latter also applies to the former.

\subsection{Correct Stabilization}
We say that a configuration is {\em stable} if and only if no process can execute a move in 
this configuration. 
We now proceed to show that if Algorithm \ref{alg:matching} reaches a stable configuration then
the $p$ and $m$-values will define a maximal matching $M$ where $(i,j)\in M$ if and only if
$(i,j) \in E, p_i = j$, and $p_j =i$ while both $m_i$ and $m_j$ are true.
In order to perform the proof, we define the following five mutual exclusive predicates:


\noindent
\begin{tabbing}
PRcondemned(i)x\= xx \=   \kill
$PRmarried(i)$   \> $\equiv$ \> $\exists j \in N(i) : (p_i = j$ {\bf and} $p_j = i)$ \\
$PRwaiting(i)$   \> $\equiv$ \> $\exists j \in N(i) : (p_i = j$ {\bf and} $p_j \not = i$ {\bf
and} $\neg PRmarried(j))$ \\
$PRcondemned(i)$ \> $\equiv$ \> $\exists j \in N(i) : (p_i = j$ {\bf and} $p_j \not = i$ {\bf and} $PRmarried(j))$\\
$PRdead(i)$      \> $\equiv$ \> $(p_i = null)$ {\bf and} $(\forall j \in N(i) : PRmarried(j))$ \\
$PRfree(i)$      \> $\equiv$ \> $(p_i = null)$ {\bf and} $(\exists j \in N(i) : \neg PRmarried(j))$
\end{tabbing}
Note first that each process will evaluate exactly one of these predicates to true.
Moreover, also note that {\em PRmarried(i)} is the same as in Algorithm \ref{alg:matching}.

We now show that in a stable configuration each process $i$
evaluates either {\em PRmarried(i)} or {\em PRdead(i)} to true, and when this
is the case, the $p$-values define a maximal matching.
To do so, we first note that in any stable configuration the $m$-values reflects the current
status of the process.

\begin{lemma}
\label{lem:mi}
In a stable configuration we have $m_i = PRmarried(i)$ for each $i \in V$.
\end{lemma}
\begin{proof}
This follows directly since if $m_i \not = PRmarried(i)$ then $i$ is eligible to
execute the {\em Update(i)} rule.
\end{proof}

We next show in the following three lemmas that no process will evaluate 
either Predicate $PRwaiting(i)$, $PRcondemned(i)$, or $PRfree(i)$ to true in a stable configuration.

\begin{lemma}
\label{lem:cond}
In a stable configuration $PRcondemned(i)$ is false for each $i \in V$.
\end{lemma}
\begin{proof}
If there exists at least one process $i$ in the current configuration
such that Predicate $PRcondemned(i)$ is true
then $p_i$ is pointing to a process $j \in N(i)$ that is married to 
a process $k$ where $k \not = i$. From Lemma \ref{lem:mi} it follows that
in a stable configuration we have $m_i = PRmarried(i)$ and $m_j = PRmarried(j)$.
Thus in a stable configuration the predicate 
$(m_i = PRmarried(i)$ {\bf and} $p_i=j$ {\bf and} $p_j \not = i$ {\bf and} $m_j)$ 
evaluates to true. But then process $i$ is eligible to execute the $Abandonment$
rule contradicting that the current configuration is stable.
\end{proof}

\begin{lemma}
\label{lem:wait}
In a stable configuration $PRwaiting(i)$ is false for each $i \in V$.
\end{lemma}
\begin{proof}
Assume that the current configuration is stable and that
there exists at least one process $i$ such that $PRwaiting(i)$ is true.
Then it follows that $p_i$ is pointing to a process $j \in N(i)$ such that
$p_j \not = i$ and $j$ is unmarried. Note first that if $p_j = null$
then process $j$ is eligible to execute a {\em Marriage} move. Also, if 
$j<i$ then process $i$ can execute an {\em Abandonment} move.

Assume therefore that $p_j \not = null$ and that $j>i$. It then follows from Lemma \ref{lem:cond}
that\linebreak $\neg PRcondemned(j)$ is true and since $j$ is not married we also
have \linebreak $\neg PRmarried(j)$. Thus $PRwaiting(j)$ must be true.
Repeating the same argument for $j$ as we just did for $i$ 
it follows that if both $i$ and $j$ are ineligible for a move then
there must exist a process $k$ such that $p_j = k$, $k>j$, and $PRwaiting(k)$ also evaluates
to true. This sequence of processes cannot be extended indefinitely since each process
must have a higher id than the preceding one.  Thus there
must exist some process in $V$ that is eligible for a move and the assumption
that the current configuration is stable is incorrect.
\end{proof}

\begin{lemma}
\label{lem:free}
In a stable configuration $PRfree(i)$ is false for each $i \in V$.
\end{lemma}
\begin{proof}
Assume that the current configuration is stable and that
there exists at least one process $i$ such that {\em PRfree(i)} is true.
Then it follows that $p_i=null$ and that there exists at least
one process $j \in N(i)$ such that $j$ is not married. 

Next, we look at the value of the different predicates for the process $j$.
Since $j$ is not married it follows that $PRmarried(j)$ evaluates to false.
Also, from lemmas \ref{lem:cond} and \ref{lem:wait} we have that both
$PRwaiting(j)$ and $PRcondemned(j)$ must evaluate to false. Finally, since $i$ is
not married we cannot have $PRdead(j)$. Thus we must have $PRfree(j)$.
But then the process with the smaller id of $i$ and $j$ is eligible to
propose to the other, contradicting the fact that the current configuration is
stable.
\end{proof}

From lemmas \ref{lem:cond} through \ref{lem:free} we immediately get the following corollary.
\begin{corollary}
\label{cor:stable}
In a stable configuration either $PRmarried(i)$ or  $PRdead(i)$ holds for
every $i \in V$.
\end{corollary}

We can now show that a stable configuration also defines a maximal matching.

\begin{theorem}
\label{the:terminal}
In any stable configuration the $m$ and $p$-values define a maximal matching. 
\end{theorem}
\begin{proof}
From Corollary \ref{cor:stable} we know that either $PRmarried(i)$ or $PRdead(i)$ holds for
every $i \in V$ in a stable configuration.  
Also, from Lemma \ref{lem:mi} it follows that $m_i$ is true if and only if $i$ is married.
It is then straightforward to see that the $p$-values define a matching. 

To see that this matching is maximal assume to the contrary that it is
possible to add one more edge $(i,j)$ to the matching so that it still
remains a legal matching. To be able to do so we must have $p_i =
null$ and $p_j =null$. Thus we have $\neg PRmarried(i)$ and $\neg
PRmarried(j)$ which again implies that both $PRdead(i)$ and
$PRdead(j)$ evaluates to true. But according to the $PRdead$ predicate
two adjacent processes cannot be dead at the same time.  It follows
that the current matching is maximal.
\end{proof}

\subsection{Convergence for the Distributed Adversarial Daemon}
\label{sub:dist}

In the following we will show that Algorithm \ref{alg:matching} will reach a stable configuration
after at most $3 \cdot n + 2 \cdot m$ steps under the distributed adversarial daemon.

First we note that as soon as two processes are married they will remain so for the rest
of the execution of the algorithm.

\begin{lemma}
\label{lem:remains}
If processes $i$ and $j$ are married in a configuration $C$ ($p_i = j$ and $p_j = i$) then they
will remain married in any ensuing configuration $C'$.
\end{lemma}
\begin{proof}
Assume that $p_i=j$ and $p_j=i$ in some configuration $C$.  Then process $i$ cannot execute
neither the {\em Marriage} nor the {\em Seduction} rule since these require 
that $p_i = null$. Similarly, $i$ cannot execute the 
{\em Abandonment} rule since this requires that $p_j\neq i$. 
The exact same argument for process $j$ 
shows that $j$ also cannot execute any of the three rules {\em Marriage}, {\em Seduction}, and 
{\em Abandonment}.
Thus the only rule that processes $i$ and $j$ can execute is {\em Update} 
but this will not change the values of $p_i$ or $p_j$. 

\end{proof}

A process discovers that it is married through executing the {\em Update} rule. Thus this is
the last rule a married process will execute in the algorithm. This is reflected in 
the following.

\begin{corollary}
\label{cor:mi}
If a process $i$ executes an {\em Update} move and sets $m_i =$ true then $i$ will not
move again.
\end{corollary}
\begin{proof}
From the predicate of the {\em Update} rule it follows that when process $i$ sets $m_i =$ true
there must exist a process $j \in N(i)$ such that $p_i=j$ and $p_j=i$. Thus from Lemma
\ref{lem:remains} the only move $i$ can make is an {\em Update} move. But since the $m_i$ value
is correct and $p_i$ and $p_j$ will not change again this will not happen.

\end{proof}

Since a married process cannot become ``unmarried''
we also have the following restriction on the number of times the  
{\em Update} rule can be executed by any process.

\begin{corollary} \label{cor:update}
Any process executes at most two {\em Update} moves.
\end{corollary}

We will now bound the number of moves from the set \linebreak $\{${\em Marriage}, {\em Seduction}, 
{\em Abandonment}$\}$. 
Each such move is performed by a process $i$ in relation to one of its neighbors $j$. 
We will call any such move made by either $i$ or $j$ with respect to the other as an $i,j$-{\em move}.

\begin{lemma}
\label{lem:time}
For any edge $(i,j) \in E$, there can at most be three steps in which an $i,j$-move is performed.
\end{lemma}
\begin{proof}
Let $(i,j) \in E$ be an edge such that $i < j$.
We then consider four different cases depending on the initial values of $p_i$ and $p_j$ 
at the start of the algorithm. For each case we will show that there can at most be 
three steps in which $i,j$-moves occur. \\

{\em Case (i):} $p_i \not = j$ and $p_j \not = i$.
Since $i < j$ the first $i,j$-move cannot be process $j$ executing a {\em Seduction} move. 
Also, as long as $p_i \not = j$,
process $j$ cannot execute a {\em Marriage} move. Thus process $j$ cannot execute an $i,j$-move
until after process $i$ has first made an $i,j$-move. It follows that the first possible $i,j$-move
is that $i$ executes a {\em Seduction} move simultaneously as $j$ makes no move. 
Note that at the starting configuration of this move we must have $\neg m_j$.
If the next $i,j$-move is performed by $j$ simultaneously as $i$ performs no move
then this must be a {\em Marriage} move which results in $p_i = j$ and $p_j = i$. Then by Lemma
\ref{lem:remains} there will be no more $i,j$-moves. If process $i$ makes the next $i,j$-move 
(independently of what process $j$ does) then this must
be an {\em Abandonment} move. But this requires that the value of $m_j$ has changed from false to true.
Then by Corollary \ref{cor:mi} process $j$ will not make any more $i,j$-moves and since $p_j \not = null$
and $p_j \not = i$ for the rest of the algorithm it follows that process $i$ cannot execute any 
future $i,j$-move. Thus there can at most be two steps in which $i,j$-moves are performed. \\

{\em Case (ii):} $p_i = j$ and $p_j \not = i$.
If the first $i,j$-move only involves process $j$ then this must be a {\em Marriage} move resulting in
$p_i = j$ and $p_j = i$ and from Lemma \ref{lem:remains} neither $i$ nor $j$ will make any future 
$i,j$-moves.
If the first $i,j$-move involves process $i$ then it must make an {\em Abandonment} move. 
Thus in the configuration prior to this move we must have $m_j=$ true. It follows that either 
$m_j \not = PRmarried(j)$ or $p_j \not = null$. In both cases process $j$ 
cannot make an $i,j$-move simultaneously as $i$ makes its move.
Thus following the {\em Abandonment} move by process $i$ we are at  Case {\em (i)}
and there can at most be two more $i,j$-moves. Hence, there can at most
be a total of three steps with $i,j$-moves. \\

{\em Case (iii):} $p_i \not = j$ and $p_j = i$.
If the first $i,j$-move only involves process $i$ then this must be a {\em Marriage} move resulting in
$p_i = j$ and $p_j = i$ and from Lemma \ref{lem:remains} neither $i$ nor $j$ will make any future 
$i,j$-moves.
If the first $i,j$-move involves process $j$ then this must be an {\em Abandonment} move. 
If process $i$ does not make a simultaneous $i,j$-move then this will result in configuration 
$i)$ and there can at most be two more steps with $i,j$-moves for a total of three steps containing
$i,j$-moves. 

If process $i$ does make a simultaneous $i,j$-move then this must be a {\em Marriage} move.
We are now at a similar configuration as Case {\em (ii)} but with $\neg m_j$. If the second $i,j$-move
involves process $i$ then this must be an {\em Abandonment} move implying that $m_j$ has changed
to true. It then follows from Corollary \ref{cor:mi} that process $j$ (and
therefore also process $i$) will not make
any future $i,j$-move leaving a total of two steps containing $i,j$-moves. 
If the second $i,j$-move does not involve $i$ then this must be a {\em Marriage} move
performed by process $j$ and resulting in $p_i=j$ and $p_j=i$ and from Lemma \ref{lem:remains}
neither $i$ nor $j$ will make any future $i,j$-moves. \\

{\em Case (iv):} $p_i = j$ and $p_j = i$.
In this case it follows from Lemma \ref{lem:remains} that neither process $i$ nor process $j$ will make 
any future $i,j$-moves.

\end{proof}

It should be noted in the proof of Lemma \ref{lem:time} that only an edge $(i,j)$ where we
initially have either $p_i = j$ or $p_j = i$ (but not both) can result in three $i,j$-moves, otherwise
the limit is two $i,j$-moves per edge. Thus there is at most one edge incident on each process
that can result in three $i,j$-moves. From this observation we can now give the following bound on
the total number of steps needed to obtain a stable solution.

\begin{theorem}
\label{the:n+m}
Algorithm \ref{alg:matching} will stabilize after at most $3 \cdot n + 2 \cdot m$ steps under the 
distributed adversarial daemon.
\end{theorem}
\begin{proof}
From Corollary \ref{cor:mi} we know that there can be at most $2n$ {\em Update} moves, each which can
occur in a separate step. From Lemma \ref{lem:time} it follows that there can at most
be three $i,j$-moves per edge. But as observed, there is at most one edge incident on each
process for which this can occur, otherwise the limit is two $i,j$-moves. Thus the total number
of $i,j$-moves is at most $n + 2 \cdot m$ and the result follows.

\end{proof}

From Theorem \ref{the:n+m} it follows that Algorithm \ref{alg:matching} will use $O(m)$ moves on
any connected system when assuming a distributed daemon. Since the distributed daemon encompasses
the sequential daemon this result also holds for the sequential daemon.

\subsection{Convergence for the Distributed Fair Daemon}
Next we consider the number of rounds used by Algorithm \ref{alg:matching} when operated
under the distributed fair daemon. 
Note that one round may encompass several steps, and we only require that every process eligible at the start of a round
either executes at least one rule during the round or becomes ineligible to do so.
This also implies that moves made in the same round may or may not be simultaneous.
Since the fair distributed daemon is a subset of the adversarial distributed daemon any results
that were shown in Section \ref{sub:dist} also applies here.
We will now show that Algorithm \ref{alg:matching} converges after at most $2 \cdot n + 1$ rounds
for this daemon.

We define that a process $i \in V$ is {\em active} if either $PRmarried(i)$ or 
$PRdead(i)$ is false.  A process that is not active is {\em inactive}.
From Corollary \ref{cor:mi} it follows that any process $i \in V$ where $PRmarried(i)$
is true will not become active again for the rest of the algorithm. This also implies
that if $PRdead(i)$ is true in some configuration then it will remain so for the
rest of the algorithm.

\begin{lemma}
\label{lem:sync}
Let $A \subseteq V$ be a maximal connected set of active processes in some configuration
of the algorithm.  If $|A|>2$ then 
after at most four more rounds the size of $A$ has decreased by at least $2$.
\end{lemma}

\begin{proof}
We first note that the size of $A$ cannot increase during the execution of
the algorithm.
Assume now that no processes in $A$ gets married during the next four rounds.
We will show that this leads to a contradiction.

After the first round every process $j \in N(A)$
must have $m_i = true$. This follows since any process $j\in N(A)$ must have
$PRdead(j) =$ false (by definition) and will therefore have $PRmarried(j) =$ true. 
Thus if $m_j$ is initially false for a process  $j \in N(A)$ then after the first round 
$m_j$ will be set to true. Similarly, if a node $i \in A$ has $m_i = $ true then $m_i$
will be set to false after the first round. According to the assumption that
no processes in $A$ gets married, the $m$-values will not change during the next three 
rounds.

Next, consider any $i \in A$ that either initially or after the first round 
satisfies $p_i = j$ such that either $j \in N(A)$ or $j<i$ (or both). 
It follows that if $j \in N(A)$ then $m_j =$ true after the first round,
and if $j<i$ then $i$ will be eligible for an {\em Abandonment} move before
$j$ can execute a {\em Marriage} move (otherwise they get married). 
Thus in either case,
process $i$ is eligible for an {\em Abandonment} move no later than after
the first round. Also note that the situation where  $p_i = j$ and
$j<i$ cannot occur again after the first round. This is because prior
to this configuration we must have $p_j = i$ and $m_i = $ true, which is
not possible if $i \in A$.

Thus after the second round 
a process $i \in A$ cannot execute an {\em Abandonment} move since this requires
that either $m_{p_i} = $ true or that $i>p_i$. Since no process can execute
an {\em Abandonment} move it also follows that no process can execute a {\em Marriage} move 
since this would lead to two processes getting married.
Thus at this stage a process can only execute a {\em Seduction} move and a process that
is not eligible for a {\em Seduction} move at this point will not become eligible
for a {\em Seduction} move
after the third round since no $m$-value is changed and no $p$-value is
set to $null$ during the third round.

Hence, at the start of the third round we have that for 
every $i \in A$ either {\em (i)} $p_i = null$ or 
{\em (ii)} $p_i = j$ where $j \in N(j) \cap A$. If Case {\em (i)} is true for 
every process in $A$, then since $|A|\geq 2$ then at least the process 
with the lowest id in $A$ is eligible for a {\em Seduction} move. 
Therefore no later than after the third round there exists at least one process 
$i_1 \in A$ where $p_{i_1} = i_2$ such that $i_2 \in N(j) \cap A$. 
Further, let $\{i_1, i_2, ..., i_k\}$ be a path of maximal length 
such that $i_{x+1} \in N(i_x) \cap A$ and $p_{i_x} = 
i_{x+1}$, $1 \leq x < k $. Note that while the \emph{Seduction} moves made by the processes during the third round may be performed in different steps, no process will become eligible for an \emph{Update} or \emph{Abandonment} move, since they must be preceded by a \emph{Marriage} and \emph{Update} move, respectively.
It follows that each $i_x \in A$ and also that $i_x < i_{x+1}$.  
Since the length of the path is finite we have $p_{i_k} = null$. 

The process $i_k$ is now eligible for a {\em Marriage} move and therefore 
cannot be eligible for any other move.
As noted, process $p_{i_{k-1}}$ cannot be eligible
for an {\em Abandonment} move at this point since $i_{k-1} < i_k$ and $m_k =$ false.
Thus following the fourth round processes $i_{k-1}$ and $i_k$ will become married,
contradicting our assumption and the result follows. 
\end{proof}

Note that if $A$ in Lemma \ref{lem:sync} only contains one node $i$ then either
$PRwaiting(i)$ or $PRcondemned(i)$ must be true initially. In either case,
after at most two moves $i$ will have updated $m_i$ and executed an \linebreak {\em Abandonment}
move such that $PRdead(i)$ is true.

Obviously $|A| \le |V|$, and from Lemma \ref{lem:remains} we know that 
once married, a process will remain married for the rest of the algorithm.
From this we get that at 
most $2 \cdot n$ rounds are needed to find the matching. However, 
after the matching has been found every married process may execute an 
{\em Update} move, and every unmarried process may execute an 
{\em Abandonment} move. Both of these can be done in the same round. 
Note that it is not necessary for a process $i$ that is unmarried when the algorithm
terminates to execute a final {\em Update} move as $m_i = $ false after the first
round and remains false throughout the algorithm. 
From this we get the following theorem.

\begin{theorem}
Algorithm \ref{alg:matching} will stabilize after at most $2 \cdot n + 1$ rounds when 
using a fair distributed daemon.
\end{theorem}

\section{Conclusion}
\label{sec:conclusion}

We have presented a new self-stabilizing algorithm for the maximal matching problem 
that improves the time step complexity 
of the previous best algorithm for the distributed adversarial daemon, 
while at the same time as meeting the bounds of the previous best algorithms for the sequential and the distributed fair
daemon.

It is well known that a maximal matching is a $\frac{1}{2}$-approximation to the {\em maximum} matching, where 
the maximum matching is a matching such that no other matching with strictly greater size exists in the network. 
In~\cite{GHS06}, Goddard \emph{et al.} provide a $\frac{2}{3}$-approximation for a particular class of networks (trees and
rings of size not divisible by $3$). Also, in particular networks such as Trees in~\cite{KS00,BM03} or bipartite graphs in ~\cite{CHS02}, self-stabilizing algorithms have been proposed for maximum matching. However, no self-stabilizing solution with a better approximation ratio than $\frac{1}{2}$ currently exists for general graphs. Thus it would be of interest to know if it is possible to create a self-stabilizing algorithm for general graphs that achieves a better approximation ratio than $\frac{1}{2}$, or even an optimal solution.

\bibliographystyle{plain}
\bibliography{matching}

\end{document}